\documentclass[preprint,prd,nofootinbib,tightenlines,amsmath,amssymb]{revtex4}
\pdfoutput=1
\usepackage[dvipdfmx]{graphicx}
\usepackage{bm}
\usepackage{amsmath}
\usepackage{hhline}
\usepackage{color}
\usepackage{xcolor}
\usepackage{slashed}
\usepackage{relsize}

\newcommand{\Hinf}{H_\text{inf}}

\begin{document}

\vspace*{3em}

\preprint{TU-1080}%
\preprint{IPMU18-0210}%

\title{Relaxing the Cosmological Moduli Problem \\
by Low-Scale Inflation} 

\author{
Shu-Yu Ho,$^{1}$\footnote[1]{ho.shu-yu.q5@dc.tohoku.ac.jp}
Fuminobu Takahashi,$^{1,2}$\footnote[2]{fumi@tohoku.ac.jp}
and 
Wen Yin$^{3}$\footnote[3]{yinwen@kaist.ac.kr}}
\affiliation{
${}^{1}$Department of Physics, Tohoku University, Sendai, Miyagi 980-8578, Japan \vspace{3pt} \\
${}^{2}$Kavli Institute for the Physics and Mathematics of the Universe (Kavli IPMU), UTIAS, WPI, The University of Tokyo, Kashiwa, Chiba 277-8568, Japan
 \vspace{3pt} \\
${}^{3}$ Department of Physics, Korea Advanced Institute of Science and Technology, Daejeon 34141, Korea \vspace{3ex} 
}

\date{\today $\vphantom{\bigg|_{\bigg|}^|}$}

\begin{abstract}
We show that the cosmological abundance of string axions is much smaller than naive estimates if the Hubble scale of 
inflation, $H_{\rm inf}$, 
is sufficiently low (but can still be much higher than the axion masses) and if the inflation lasts sufficiently long. The reason is that the initial misalignment angles of the string axions follow the Bunch-Davies distribution peaked at the potential minima. As a result, the cosmological moduli problem induced by the string axions can be significantly relaxed by low-scale inflation, and astrophysical and cosmological bounds are satisfied 
over a wide range of the mass without any fine-tuning of the initial misalignment angles. Specifically, the axion with its decay constant $f_\phi = 10^{16}$\,GeV satisfies the bounds over $10^{-18}{\rm \, eV} \lesssim m_\phi \lesssim 10{\rm\,TeV}$ for  $H_{\rm inf} \lesssim 10{\rm\,keV}- 10^{6}$\,{\rm GeV}. We also discuss cases with multiple axions and the QCD axion. 
\end{abstract}

\maketitle

\section{Introduction}\label{sec:1}
Light axions may be ubiquitous in nature. In string theory, there often appear (relatively) light scalar moduli through compactification~\cite{Polchinski:1998rq}. If supersymmetry (SUSY) survives below the compactification scale, a modulus forms a chiral supermultiplet, $X$. Its lowest component can be decomposed as $X = r + i  \phi$, where $r$ and $\phi$ denote the real and imaginary parts, respectively. We identify $\phi$ with an axion which enjoys discrete shift symmetry,
\begin{eqnarray}\
\label{shift}
\phi  \rightarrow \phi + 2  \pi f_\phi ~,
\end{eqnarray}
where $f_\phi$ is the decay constant of the axion. 

While some of the moduli may drive inflation in the early Universe, explain the current accelerated cosmic expansion or become (a part of) dark matter (DM), others can cause serious cosmological problems~\cite{Coughlan:1983ci,deCarlos:1993wie}. The cosmological impact depends on
the moduli masses fixed by the moduli stabilization mechanism. While most of the moduli fields are known to be stabilized by fluxes~\cite{Grana:2005jc,Blumenhagen:2006ci},  some of them remain light, and they are stabilized by non-perturbative and/or SUSY breaking effects~\cite{Gorlich:2004qm}.  For instance, in the KKLT mechanism~\cite{Kachru:2003aw}, the  K\"ahler modulus is stabilized by instantons/gaugino condensations, and it acquires a SUSY mass parametrically larger than the gravitino mass. In particular, both real and imaginary components have almost the same mass. On the other hand, it is possible that the real components of the moduli fields are stabilized by SUSY breaking effects, while their axionic partners remain light due to shift symmetry~\cite{Svrcek:2006yi,Conlon:2006tq,Choi:2006za,Arvanitaki:2009fg,Acharya:2010zx,Higaki:2011me,Cicoli:2012sz,Cicoli:2013rwa,Visinelli:2018utg}.
In this case, the real components typically have a mass of order the gravitino mass, but the axions acquire much lighter masses from non-perturbative effects in the low energy. In particular, the axion masses may be spread over many orders of magnitude known as the Axiverse~\cite{Arvanitaki:2009fg}. 
We will focus on the latter case where the axions remain light in the low energy while the 
 real components are stabilized by the SUSY breaking effects, for reasons that will become clear shortly. 

In general, a light scalar field can be copiously produced after inflation when its starts to oscillate about the 
potential minimum. This is because, if its mass is lighter than the Hubble parameter during inflation, $H_{\rm inf}$,
 the initial position of the scalar field is generically deviated from the low-energy 
potential minimum. On the other hand, a heavy scalar with 
its mass much larger than $H_{\rm inf}$ is already stabilized at the potential minimum during inflation, and therefore, its cosmological
abundance is negligibly small. 
 In the following, we focus on the cosmological abundance of the lightest axion field, because it is more likely produced by the
 above mechanism than the heavier scalars. 
 We will come back to the case with multiple axion fields later in this paper. The masses of the real components tend to be heavier than
 their axionic partners, and their cosmological abundance can be greatly suppressed if their masses are larger than $H_{\rm inf}$ which we assume throughout this paper.

During inflation, the initial position of the axion field is considered to be deviated from the potential minimum by a factor 
of the decay constant $f_\phi$. After inflation ends, the Hubble parameter starts to decrease. 
When the Hubble parameter becomes comparable to the axion mass,
the axion starts to oscillate around the potential minimum with a large initial amplitude of order $f_\phi$.
 For the decay constant $f_\phi$ of ${\cal O}(10^{16})$\,GeV, the axion abundance is so large that it comes to dominate the Universe soon after the reheating, causing various cosmological problems. 
If the axion is sufficiently light, it may be stable on a cosmological time-scale. Then the axion abundance often exceeds the observed DM abundance by many orders of magnitude. If the axion is unstable and decays into the standard model (SM) particles such as a pair of photons, then,  its decay products may change the light element abundances,
spoiling the success of big bang nucleosynthesis (BBN)~\cite{Kawasaki:2004qu,Kawasaki:2017bqm}, or produce too much X-ray or gamma-ray fluxes~\cite{Kawasaki:1997ah,Asaka:1997rv, Asaka:1999xd}. Even if the axion decays well before the BBN, its decay releases a large entropy that dilutes any pre-existing baryon asymmetry. Furthermore, in general, a heavy scalar decay may overproduce unwanted relics which cause similar cosmological problems~\cite{Endo:2006zj,Nakamura:2006uc,Dine:2006ii,Endo:2006tf,Cicoli:2012aq,Higaki:2012ar,Higaki:2013lra}.

The axion abundance can be suppressed in various ways.
For instance, thermal inflation is known to produce large entropy to dilute the axion abundance~\cite{Yamamoto:1985rd,Lyth:1995ka}. A potential problem of this solution is that any pre-existing baryon asymmetry is similarly diluted, and one needs to invoke either an efficient baryogenesis mechanism~\cite{Affleck:1984fy,Dine:1995kz,Kasuya:2001tp}
or late-time baryogenesis~\cite{Stewart:1996ai}. 
Another simple possibility somewhat similar to thermal inflation is to suppose that $\Hinf$ is smaller than the scalar mass~\cite{Randall:1994fr}. In this case, the axion is already stabilized at the potential minimum during inflation (as we assume for the real component), and therefore, 
its abundance is significantly suppressed. However, if the axion is very light, this solution requires rather low-scale inflation for which successful reheating as well as inflation model building itself might be far from trivial to achieve (see, e.g., Refs.~\cite{Daido:2017wwb, Daido:2017tbr, Takahashi:2019qmh}).
See Refs.~\cite{Linde:1996cx,Nakayama:2011wqa,Nakayama:2011zy} for another solution to the moduli problem.

In this paper, we show that the cosmological abundance of string axions can be significantly suppressed if the inflation scale $H_{\rm inf}$ 
is low but still higher than the axion masses,\footnote{
In general,  low-scale inflation involves small parameters. While the low inflation scale may be realized by non-perturbative effects through
dimensional transmutation without fine-tuning,  the initial condition of the inflaton must be carefully chosen near the flat plateau of the potential
where the slow-roll inflation is possible. In our scenario, we further assume a very long duration of the inflation, and this is possible if 
the inflaton potential allows eternal inflation~\cite{steinhardt-nuffield,Vilenkin:1983xq,Linde:1986fc,Linde:1986fd,Goncharov:1987ir,Guth:2007ng}. For instance, hilltop inflation can do the job. In the context of eternal inflation, it is far from trivial
to quantify the amount of fine-tuning of the parameters, because the required fine-tuning of the initial condition 
might be partially canceled by the exponential expansion of the Universe (see also e.g. Refs.~\cite{Guth:2000ka,
Vilenkin:2006xv,Winitzki:2006rn,Linde:2007fr} for reviews for measure problem).} and if the inflation lasts sufficiently long. This is because the probability distribution of the axion field reaches equilibrium known as the Bunch-Davies (BD) distribution~\cite{Bunch:1978yq} where the dissipation due to quantum fluctuations is balanced by the classical motion. Interestingly, even though the Hubble parameter is much larger than the axion mass, the probability distribution of the axion field is still peaked at the potential minimum. In other words, the axion knows the location of the minimum 
 in a probabilistic way. Therefore, the axion abundance turns out to be much smaller than naive estimates, since the typical value of the initial amplitude can be significantly suppressed. We note that a similar mechanism was recently applied to the QCD axion and it was shown that the QCD axion window is open up to the Planck scale if the inflation scale is lower than the QCD confinement scale~\cite{Graham:2018jyp,Guth:2018hsa}. Here we show that the mechanism also works for string axions.

Lastly, let us mention an important requirement for the above mechanism using the BD distribution to work. Since the axion starts to oscillate around the minimum after inflation ends, the potential minimum during inflation should almost coincide with the low-energy minimum, since otherwise the BD distribution is peaked at a wrong place and the initial oscillation amplitude is not suppressed. To this end, we focus on an imaginary part of the modulus field, $\phi$. This is because the axion
potential arises from some non-perturbative effects and it is plausible that the axion potential remains unchanged during and after inflation. On the other hand, the potential of the real component, $r$, is generically modified by the SUSY breaking effect during inflation, and there is no special reason to expect that the potential minimum during inflation coincides with that after inflation unless $H_{\rm inf}$ is much smaller than the mass of $r$. Moreover, if the mass of $r$ is heavy enough to decay well before the BBN, its cosmological impact will be much milder.
In Ref.~\cite{Randall:1994fr}, it was briefly commented that the BD distribution may solve the moduli problem induced by $X$ (or $r$) 
with the mass of order the gravitino mass, but this possibility
was disregarded because the required duration of inflation was considered too long. In fact, such a long period of inflation can be realized with eternal inflation. In addition, as noted above,  this solution applies only to the case where the position of the minimum does not change after inflation, which is a plausible assumption for the string axions, not the real components.

The structure of this paper is organized as follows. In the next section, we briefly review the cosmological moduli problem and the current astrophysical and cosmological bounds on the modulus abundance. In Sec.~\ref{sec:3}, we study how the axion abundance can be suppressed by the low-scale inflation. The last section is devoted to discussion and conclusions. 

%%%%%%%%%%%%%%%%%%%%%%%%%%%%%%%%%%%%%%%%%%%%%%%%%%%%%%%%%%%%%%%%

\section{Cosmological Moduli Problem}\label{sec:2} 
\subsection{Cosmological abundance}
We focus on an imaginary component $\phi$ of a light modulus $X$, assuming that the real component $r$
has a much heavier mass and therefore its cosmological impact is not as significant as $\phi$. This is the case if $r$ decays much before
the BBN starts or if it is so heavy that its cosmological abundance is negligible. 

The potential of the axion is generated by non-perturbative effects. Due to the discrete shift symmetry (\ref{shift}),
it is periodic with period $2\pi f_\phi$. In the simplest case, the potential is given by a cosine term,
\begin{eqnarray}\label{pot}
V(\phi) 
\,=\, \Lambda^4
\bigg[
1-\cos\bigg(\frac{\phi}{f_\phi}\bigg)
\bigg]
\,\simeq\, \frac{1}{2}  m_{\phi}^2  \phi^2
~,
\end{eqnarray}
where $\Lambda \equiv \sqrt{m_\phi f_\phi}$ corresponds to the dynamical scale, and we have approximated the potential
as the quadratic one assuming $|\phi| \lesssim f_\phi$ in the second equality.

During inflation the position of the axion field generically deviates from the vacuum, $\phi = 0$, if $m_{\phi} \ll H_\text{inf}$.
This is because the axion is frozen to some field value due to the Hubble friction.\footnote{In the next section we will see how this picture is
modified by taking account of quantum fluctuations.} 
After inflation ends, the Hubble parameter starts to decrease. Then, when the Hubble 
parameter becomes comparable to the axion mass, the axion starts to oscillate around the potential minimum with an initial amplitude 
$\phi_{\rm ini}$. The energy density at the initiation of coherent oscillations is given by
\begin{eqnarray}
\rho_{\phi,{\rm ini}}  \,\simeq\,  \frac{1}{2}  m_{\phi}^2  \phi_{\rm ini}^2 ~,
\end{eqnarray}
and afterwards, the axion energy density decreases as $R(t)^{-3}$ due to the cosmic expansion with the scale factor $R(t)$.
The relic abundance of the axion coherent oscillations depends on whether the axion starts to oscillate after or before the
reheating. We will consider the two cases in the following.

First, let us consider the case in which the axion starts to oscillate during the radiation dominant era after the reheating.
This is the case if the inflaton decay rate is larger than the axion mass, $\Gamma_{\rm inf} > m_\phi$. 
In the radiation dominant epoch, the Hubble parameter is given by
\begin{eqnarray}
H(T) \,=\, \left(\frac{\pi^2 g_\star(T)}{90}\right)^{1/2} \frac{T^2}{M_\text{pl}} ~,
\end{eqnarray}
where $M_\text{pl} \,\simeq\, 2.4 \times 10^{18} \, \text{GeV}$ is the reduced Planck mass, and $g_\star(T)$ is the effective 
relativistic degrees of freedom contributing to the energy density. The axion starts to oscillate at the plasma temperature
$T = T_\text{osc}$, where $T_\text{osc}$ is given by
\begin{eqnarray}
T_\text{osc} \,\equiv\, 
\bigg(\frac{90}{\pi^2 g_{\star,\text{osc}}}\bigg)^{1/4}
\sqrt{m_\phi  M_\text{pl}}
\,\,\simeq\,
2.7 \times 10^8 \,\text{GeV} \,
\bigg(\frac{g_{\star,\text{osc}}}{106.75}\bigg)^{-1/4}
\bigg(\frac{m_\phi}{0.1\,\text{GeV}}\bigg)^{1/2} ~,
\end{eqnarray} 
where $g_{\star,\text{osc}} \equiv g_\star(T_\text{osc})$. The ratio of the axion energy density  to the entropy density 
of the Universe is then given by
\begin{eqnarray}\label{Y1}
\frac{\rho_{\phi}}{s}
\,\simeq\,
1.2 \times 10^3 \,\text{GeV} \,
\bigg(\frac{g_{\star,\text{osc}}}{106.75}\bigg)^{-1/4}
\bigg(\frac{m_\phi}{0.1\,\text{GeV}}\bigg)^{1/2}
\bigg(\frac{\phi_\text{ini}}{10^{16}\,\text{GeV}}\bigg)^2 ~,
\end{eqnarray} 
where $s = 2\pi^2 g_s(T)  T^3/45$ with $g_s(T)$ being the effective relativistic degrees of freedom contributing to the entropy, 
and here we take  $g_s(T_\text{osc}) = g_\star(T_\text{osc})$.
In Eq.\,(\ref{Y1}), we have included an additional numerical factor $\sim 2$ obtained by solving the equation of motion of the axion.
Note that the ratio $\rho_\phi/s$ is a conserved quantity if $\phi$ is stable, since both $\rho_\phi$ and $s$ scale 
as $R(t)^{-3}$ as the Universe expands.

Secondly, let us consider the case in which the axion starts to oscillate before the reheating when
the Universe is still dominated by the non-relativistic inflaton matter. In this case,
 the axion abundance is partially diluted by the inflaton decay. By assuming an instantaneous 
 conversion of the inflaton energy density $\rho_{\rm inf}$ to the radiation energy density $\rho_R$ at the reheating, the axion abundance
can be evaluated as follows\,:
\begin{eqnarray}\label{Y2}
\frac{\rho_{\phi}}{s}
\,=\,
\frac{\rho_{\phi}}{\rho_\text{inf}} \bigg|_{\text{osc}} \,
\frac{\rho_R}{s} \bigg|_{\text{RH}} 
\,\simeq\, 
9.4 \times10^{-8} \,\text{GeV} \,
\bigg(\frac{T_\text{RH}}{20\,\text{MeV}}\bigg)
\bigg(\frac{\phi_\text{ini}}{10^{16}\,\text{GeV}}\bigg)^2,
\end{eqnarray}
where `osc' and `RH' imply that the variables are evaluated at the onset of oscillations and at the reheating, respectively.
The reheating temperature $T_{\rm RH}$ is defined by 
\begin{eqnarray}
T_\text{RH} \,\equiv\, 
\bigg(\frac{90}{\pi^2 g_\star(T_\text{RH})}\bigg)^{1/4}
\sqrt{\Gamma_\text{inf}  M_\text{pl}} ~.
\end{eqnarray}
In the first equality of (\ref{Y2}), we have used the fact that $ \rho_\phi/\rho_\text{inf}$ remains constant
over time since we assume that the decay of the inflaton is negligible before the reheating, and the equation of state of the inflaton 
matter is $0$.
In the second equality of (\ref{Y2}),
we have included an extra numerical factor $\sim 2$ obtained by solving the equation of motion for the axion. 

The axion may be coupled to the SM particles and decay into lighter particles such as photons.
If the axion has a lifetime longer than the present age of the Universe, it would contribute to DM. 
On the other hand, if the axion is unstable and decays into the SM particles, the energetic decay products may
destroy the light elements such as D, ${}^3$He, and ${}^4$He synthesized by the BBN  or overproduce the X-ray or gamma-ray fluxes.
Even if the axion decays much before the BBN, it may dilute any pre-existing baryon asymmetry, or produce too much light hidden particles contributing to dark radiation or DM. 
For later use, let us express the axion abundance in terms of the density parameter, assuming that the axion is stable\,:
\begin{eqnarray}\label{phistable}
\Omega^{\rm stable}_\phi \,\equiv\, 
\frac{\rho_{\phi}}{s} \left(\frac{\rho_c}{s_0}\right)^{-1} ~,
\end{eqnarray}
where $\rho_c/s_0 \simeq 3.6 \times 10^{-9} h^{-2}$\,GeV
denotes the ratio of the critical density to the present entropy density, $h \simeq 0.67$ is the reduced
Hubble parameter, and the subscript `$0$' means that the variable is evaluated at present. 
Note that  $\rho_\phi/s$ in Eqs.~\eqref{Y1} and \eqref{Y2} represents the primordial axion abundance which does not take account of the subsequent axion decay.
Substituting Eqs.\,\eqref{Y1} and \eqref{Y2} into Eq.\,\eqref{phistable}, we obtain
\begin{eqnarray}
\Omega^{\rm stable}_{\phi}  h^2 
\,\simeq\,
\begin{cases}
\displaystyle
3.3 \times 10^{11} 
\bigg(\frac{g_{\star,\text{osc}}}{106.75}\bigg)^{-1/4}
\bigg(\frac{m_\phi}{0.1\,\text{GeV}}\bigg)^{1/2} 
\bigg(\frac{\phi_\text{ini}}{10^{16}\,\text{GeV}}\bigg)^2
& {\rm for~\,}\Gamma_\text{inf} \,>\, m_\phi
\\[0.5cm]
\displaystyle
 2.6 \times 10
\bigg(\frac{T_\text{RH}}{20\,\text{MeV}}\bigg) 
\bigg(\frac{\phi_\text{ini}}{10^{16}\,\text{GeV}}\bigg)^2
& {\rm for~\,}\Gamma_\text{inf} \,<\,  m_\phi
\end{cases} .
\label{abundance}
\end{eqnarray}
One can see that, if the initial oscillation amplitude of the axion $\phi_\text{ini}$ is around the GUT scale, the current axion density
exceeds the DM abundance, $\Omega_{\rm DM} h^2 \simeq 0.12$,  by many orders of  magnitude over a wide range of the axion mass.
As we shall see shortly, the bound on the axion density is much severer for unstable axions. Thus, the axion (or more generically, modulus) is efficiently generated by coherent oscillations in the early Universe and its large abundance causes
various cosmological problems for a wide range of the modulus mass. 
This is the so-called cosmological moduli problem.

\subsection{Astrophysical and cosmological constraints}\label{subsec:2} 
Here let us summarize the main astrophysical and cosmological constraints on the axion abundance. 
For simplicity,  we assume that the axion  decays into two photons through the Lagrangian
\begin{eqnarray}
{\cal L} \,=\, \frac{\alpha}{4\pi}\frac{\phi}{f_\phi} F_{\mu\nu} \widetilde{F}^{\mu\nu} ~,
\end{eqnarray}
where $\alpha$ is the fine structure constant, $F_{\mu\nu}$ is the electromagnetic field strength tensor, and $\widetilde{F}_{\mu\nu}$ is its dual tensor. It follows that the axion decays into two photons at a rate 
\begin{eqnarray}
\Gamma_{\phi \to \gamma\gamma} 
\,=\, 
\frac{\alpha^2}{64\pi^3}\frac{m_\phi^3}{f_\phi^2} ~,
\end{eqnarray}
which leads to the lifetime of the axion 
\begin{eqnarray}\label{tauphi}
\tau_\phi \,=\, 
\frac{1}{\Gamma_{\phi \to \gamma\gamma}} \,\simeq\,
2.5 \times 10^{18} \, \text{sec} 
\bigg(\frac{f_\phi}{10^{16}\,\text{GeV}}\bigg)^2
\bigg(\frac{m_\phi}{0.1\,\text{GeV}}\bigg)^{-3}  ~.
\end{eqnarray} 
Therefore, the current axion abundance is related to the primordial axion abundance through
\begin{eqnarray}
\Omega_{\phi}\,=\,\Omega_{\phi}^{\rm stable} e^{-t_0/\tau_\phi} ~,
\end{eqnarray}
where $t_0 \,\simeq\, 4.4\times 10^{17}\,$sec is the present age of the Universe. In the following, we 
will take $f_\phi=10^{16}\,$GeV unless otherwise stated.

First, if the axion mass is lighter than about $0.1\,\text{GeV}$, the axion is stable on a cosmological time scale and its abundance
should not exceed the DM abundance, 
\begin{eqnarray}\label{DMbound}
\Omega_\phi h^2 \,\lesssim\,\Omega_{\rm DM} h^2 \,\simeq\, 0.12~.
\end{eqnarray}
Secondly, even if the axion lifetime is longer than the present age of the Universe, its decay produces 
diffuse photon background. Specifically, for the axion lifetime longer than the time of the recombination $t_{\rm rec}\simeq 10^{13}\, \text{sec}$,
the axion abundance is tightly constrained by the X-ray and gamma-ray fluxes. 
In fact, they provide the tightest bounds for $t_{\rm rec}\lesssim  \tau_\phi \lesssim 10^{28}$\,sec, or equivalently, $10^{-4}\,\text{GeV} \lesssim m_\phi   \lesssim 10\,\text{GeV}$.

Let us estimate the Galactic and extragalactic contributions to
the diffuse photon background. If $\tau_\phi\gtrsim t_0$, the axion constitutes a fraction ${\Omega_\phi /\Omega_{\rm DM}}$ of the 
total DM in the present Universe. We assume that the fraction remains constant through the structure formation and the axion density
follows the DM density profile. 
The spectrum of the monochromatic photons produced by the decay of a single axion particle is given by
\begin{eqnarray}
{d N_\gamma \over d E}\,\equiv\, %\frac{dN_\gamma}{dE}=
2 \delta{(E-m_\phi/2)} ~.
\end{eqnarray}
Then, the differential photon flux from the axion decay in our Galaxy is given by
\begin{eqnarray}
\frac{d\Phi_{\rm Galactic}}{dE_\gamma}
\,=\,  \int_0^\infty dy \,  \frac{\Gamma_{\phi \rightarrow \gamma \gamma} }{4\pi y^2} {d N_\gamma \over d E_\gamma}   \frac{\rho_\phi({y})}{m_\phi} y^2 d\Omega 
\,=\, \frac{r_\odot}{4\pi}\frac{\rho_{\odot}}{m_{\phi}} \Gamma_{\phi \rightarrow \gamma \gamma} {d N_\gamma \over d E_\gamma}\frac{\Omega_\phi}{\Omega_{\rm DM}} J_D ~,
\end{eqnarray}
where  $r_\odot=8.5\,$kpc is the distance between the Sun and the Galactic center, $\rho_\odot= 0.3\,{\rm GeV/cm^3}$ 
is the local DM density, and $\rho_\phi(y)$ is the axion density profile in our Galaxy.
The integration in the first equality is taken over both the solid angle of the observed area and the 
line-of-sight distance, $y$.
We have defined a $J$-factor as
\begin{eqnarray}
J_D \,\equiv\,  \int_0^\infty \frac{d y}{r_\odot} \frac{\rho_{\rm DM}(y)}{\rho_\odot} d\Omega ~,
\end{eqnarray}
where $\rho_{\rm DM}(y)$ denotes the DM density profile in our Galaxy.

The extragalactic diffuse photon flux $\Phi_{\rm ex}$ comes from the axion decay in the past.
The differential flux is similarly obtained by assuming the homogeneous distribution of axion density and by taking into account the redshift~\cite{Asaka:1997rv}, which is given by
\begin{eqnarray}
{d^2\Phi_{\rm ex} \over d \Omega d {E_\gamma}} \,=\, 
{\Gamma_{\phi\rightarrow \gamma\gamma} \over 4\pi}   \int_{t_{\rm rec}}^{t_0}{dt' n_\phi(t')(1+z)^{-3} { d E_\gamma' \over d E_\gamma } {d N_\gamma\over d E'_\gamma} ~.}
\end{eqnarray}
Here $z$ is the redshift parameter, $E_\gamma$ is the observed energy of the photon, and $E'_\gamma=(1+z)E_\gamma$ is the photon energy just after the decay. $(1+z)^{-3}$ represents the dilution of the flux due to the cosmic expansion.
We have defined the number density of the axion at the cosmic time $t$ as
\begin{eqnarray}
n_\phi(t) \,\equiv\, {\Omega_\phi^{\rm stable} \rho_c\over m_\phi } (1+z)^3e^{-t /\tau_\phi} ~.
\end{eqnarray}
The redshift parameter $z$ is related to cosmic time $t$ by
\begin{eqnarray}
\frac{dt}{dz} \,=\, -\left[H_0 {\left( 1+z\right)} \sqrt{\Omega_M (1+z)^3 +\Omega_\Lambda}\,\right]^{-1}.
\end{eqnarray}
Here, $H_0$ is the present Hubble constant, and $\Omega_{M} \simeq 0.3$ 
and $\Omega_{\Lambda} \simeq 0.7$ denote the density parameter of matter and the cosmological constant, respectively. 

\begin{figure}[t!]
\begin{center}  
\includegraphics[width=145mm]{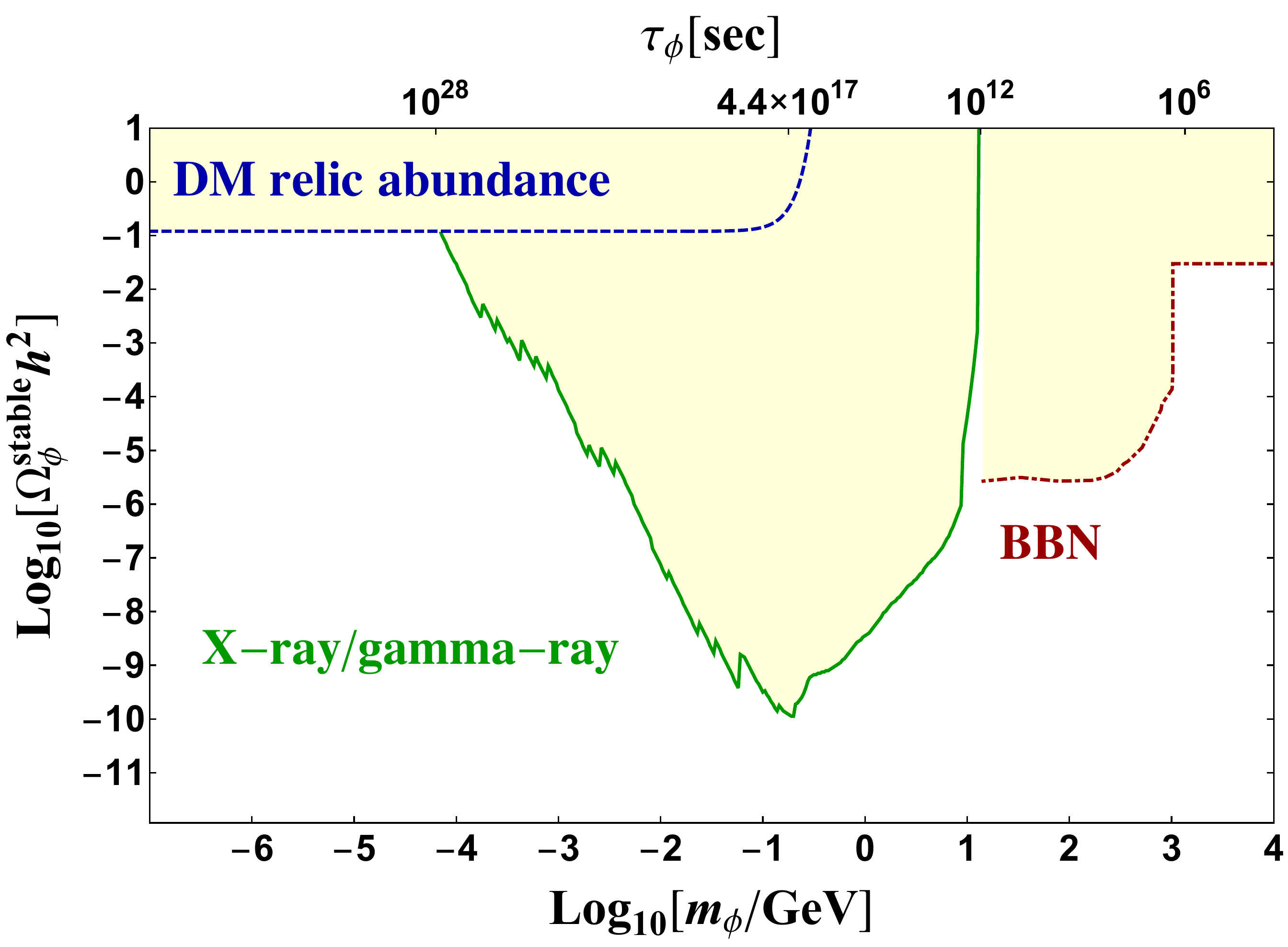}
\end{center}
\caption{The  astrophysical and cosmological bounds on the primordial axion abundance, $\Omega^{\rm stable}_\phi,$ as a function of the axion mass by fixing $f_\phi=10^{16}\,$GeV, 
where the yellow shaded region is excluded. The dashed blue line, solid green line, and
dotted-dashed red line are the constraints from the DM abundance, 
the X-ray and gamma-ray fluxes, and the BBN, respectively. The constraint from the CMB distortion is not shown here 
since this constraint is weaker than the others.} 
\label{fig:Omegavsm}
\end{figure}

Notice that since the Universe is opaque to photons at $t\lesssim t_{\rm rec}$, we have taken the cutoff of the integral to be $t=t_{\rm rec}$. 
By performing the integration, we obtain
\begin{eqnarray}\label{Gammaphi}
\frac{d^2\Phi_{\rm ex}}{d\Omega dE_\gamma} 
&=&
\frac{ \Omega^{\rm stable}_\phi \rho_c \Gamma_{\phi\to\gamma\gamma} }{2\pi m_\phi E_\gamma H_0  } 
\Bigg[ 
\Omega_{\Lambda} + \Omega_{M}\bigg(\frac{m_\phi}{2E_\gamma}\bigg)^3\,
\Bigg]^{-1/2}
\nonumber\\[0.15cm]
&&\times \exp
\Bigg[
{-}\frac{2}{3H_0\tau_\phi\sqrt{\Omega_{\Lambda}}} 
\sinh^{-1}
\Bigg(
\sqrt{\frac{\Omega_{\Lambda}}{\Omega_{M}}} 
\bigg(\frac{2E_\gamma}{m_\phi}\bigg)^{3/2}
\Bigg)
\Bigg] ~.
\end{eqnarray}
We emphasize here that this formula is valid only  for
${\frac{1}{2} m_\phi/[1+z(t_{\rm rec})] < E_\gamma < \frac{1}{2} m_\phi}$, where $z{(t_{\rm rec})}\simeq 1100$,
and there is no photon flux from axion decays outside this energy range. 
The predicted extragalactic and Galactic diffuse photon flux should be smaller than the room left for extra diffuse photon flux, which puts a tight upper bound on the energy density of the axion.

Finally, if the axion mass is above 10\,GeV or so, the axion decays shortly after or during the BBN. 
Then its decay into high energy photons may dissociate or overproduce light elements of the Universe, 
which would contradict with the primordial light element abundances inferred by observations. In order not 
to spoil the success of the BBN, the axion abundance must be sufficiently small.  
If the lifetime of the axion is longer than $10^6$\,sec and shorter than the recombination epoch,
the constraint from distortion of the CMB spectrum should also be taken into account. 
However, this constraint is not as strong as the BBN one. 

In Fig.\,\ref{fig:Omegavsm}, we show the various upper bounds on the primordial axion abundance, $\Omega^{\rm stable}_\phi h^2$,
as a function of the axion mass, $m_\phi$. 
The constraint that comes from the DM abundance \eqref{DMbound} 
applies to the axion mass below $\sim 0.1$\,GeV and extends down to  a very small mass of ${\cal O}(10^{-18})$\,eV. For even lighter axion masses, the axion cannot be a dominant component
of DM.
For the X-ray and gamma-ray limits, we have made use of the observed flux data summarized 
in Ref.~\cite{Essig:2013goa} and we assume the NFW DM density profile for $J_D$~\cite{Navarro:1995iw,Navarro:1996gj}.  We require that the predicted flux should not exceed the observed one with the $1\sigma$ error bar. 
For the BBN constraint, we have extracted several points of ${\rho_{\phi}}/s$ and $\tau_\phi$ from the analysis of the energy
injection during the BBN~\cite{Kawasaki:2017bqm}, and we made a conservative interpolation of the data points
for $m_\phi \gtrsim 10^3\,$GeV. The BBN bound becomes weak and disappear for
 $m_\phi \gtrsim {\cal O}(100)$\,TeV as its lifetime becomes much shorter than 1\,sec. As one can see from the figure and Eq.\,\eqref{abundance},
 there is a clear tension between theoretical expectation (\ref{abundance}) and observations, which necessitates some mechanism to suppress the axion abundance.

%%%%%%%%%%%%%%%%%%%%%%%%%%%%%%%%%%%%%%%%%%%%%%%%%%%%%%%%%%%%%

\section{Low-scale Inflation as a solution to the cosmological moduli problem}\label{sec:3}
\subsection{Bunch-Davies distribution}
Here we briefly review the BD distribution of a scalar field  in de Sitter Universe.
As we shall see shortly, the BD distribution is reached after a large number of $e$-folds,
which can be realized in e.g. eternal inflation~\cite{steinhardt-nuffield,Vilenkin:1983xq,Linde:1986fc,Linde:1986fd,
Goncharov:1987ir,Guth:2007ng}.
We consider a scalar field $\varphi$ with a minimal coupling to gravity given by 
\begin{eqnarray}
S \,=
\mathop{\mathlarger{\int}} d^4 x \, \sqrt{-\text{det}(g_{\mu\nu})} 
\bigg[
{-}\frac{1}{2} g^{\mu \nu} {\partial \varphi\over \partial x^\mu} {\partial \varphi\over \partial x^\nu}-V(\varphi)-V_0
\bigg]
~,
\end{eqnarray}
where $V_0 \simeq 3H_{\rm inf}^2 M_{\rm pl}^2$ is the vacuum energy, and we assume that the energy of the 
scalar field is subdominant. 
 For simplicity, we approximate the scalar potential as the quadratic one
\begin{eqnarray}
V(\varphi) \,\simeq\, 
\frac{1}{2}m_\varphi^2  \varphi^2 ~,
\end{eqnarray}
where the mass of the scalar field, $m_\varphi$, is assumed to be much smaller than the Hubble parameter during inflation, $m_\varphi \ll \Hinf$.

Let us first decompose the scalar field into a spatially homogeneous part and a fluctuation about it, 
 $\varphi({\bf x},t) = \varphi_0(t) + \delta \varphi({\bf x},t)$. In the absence of quantum fluctuations, 
 the homogeneous part $\varphi_0(t)$ will asymptote to zero as $\exp{\left[-({m_\varphi^2 / 3H_{\rm inf}})t\right]}$ due to the classical equation of motion
 after a large number of $e$-folds, $N \sim H_{\rm inf}t\gg H_\text{inf}^2/m_\varphi^2$~\cite{Dimopoulos:1988pw}. In fact, after such a large
 $e$-folding, the scalar field is dominated by (accumulated) quantum fluctuations.
The fluctuation $\delta \varphi({\bf x},t)$ can be expressed in the Fourier form as
\begin{eqnarray}
\delta\varphi({\bf x},t)
\,=
\mathop{\mathlarger{\int}}\hspace{-0.18cm} 
\frac{d^3 k}{(2\pi)^{3/2}} 
\Big[ 
\delta\varphi_k(t)  a_{\bf k}  e^{i  {\bf k \cdot x}} +
\delta\varphi^{\ast}_k(t)  a_{\bf k}^\dagger  e^{-i  {\bf k \cdot x}}
\Big]
~,
\end{eqnarray}
where $k \equiv |\bf{k}|$ denotes a comoving wavenumber. 
The coefficients $a_{\bf k}$  and $a^\dagger_{\bf k}$ are to be identified with 
the annihilation and creation operators, respectively, when quantized in a deep subhorizon regime.
At scales much smaller than the Hubble horizon,  one can neglect the effect of the gravity and canonically 
quantize the scalar field as in the Minkowski space-time. Then, one can define the BD vacuum by $a_{\bf k} | 0 \rangle = 0$ for all ${\bf k}$ with  $\left\langle {0|0} \right\rangle =1$~\cite{Bunch:1978yq}. After a sufficiently long inflation,
 the fluctuations of the scalar field on scales of order the horizon obey a Gaussian distribution (BD distribution), with a variance $\left\langle {\varphi^2}\right\rangle$ given by
\begin{eqnarray}\label{phi2}
\left\langle {\varphi^2}\right\rangle \,\simeq\, \frac{3H_\text{inf}^4}{8\pi^2m_\varphi^2} ~.
\end{eqnarray}
The typical size of $\left\langle {\varphi^2}\right\rangle$ can be understood by equating the field excursion 
by classical motion, $\Delta \varphi_{\rm classical}\sim N_{\rm eq}{m_\varphi^2\varphi/ H_{\rm inf}^2}$, 
to accumulated quantum fluctuations, $\Delta \varphi_{\rm quantum}\sim {\sqrt{N_{\rm eq}} H_{\rm inf} / ({2\pi})} 
$. {Here $N_{\rm eq}\sim H_{\rm inf}^2 /m_\varphi^2$ is the typical $e$-folding that the field excursion by the classical motion becomes important. For $N\gg N_{\rm eq}$, the $\left\langle {\varphi^2}\right\rangle$ asymptotes to Eq.\,\eqref{phi2}.} 
For more detailed derivations of Eq.\,\eqref{phi2}, see e.g. Refs.~\cite{Guth:2018hsa,Graham:2018jyp}.

\subsection{Relaxing the cosmological moduli problem by the BD distribution}

Now we identify the scalar field with the axion in the previous section, and apply the BD distribution to its
initial oscillation amplitude assuming the existence of the sufficiently long duration of inflation before
the CMB scales exited the horizon. We assume that the axion potential remains unchanged during and 
after inflation. This is considered to be the case if the
Gibbons-Hawking temperature~\cite{Gibbons:1977mu},  $T_\text{GH} = H_\text{inf}/(2 \pi)$,
 is lower than the dynamical scale $\Lambda$.

As we have seen before, if the inflation lasted sufficiently long, 
the axion field value follows the BD distribution, and its typical initial value is given by
\begin{eqnarray}\label{phi_ini}
\phi_\text{ini}^{\rm (BD)}
\,=\,
\sqrt{\left\langle {\phi^2}\right\rangle} \,=\,
\sqrt{\frac{3}{8\pi^2}}\frac{H_\text{inf}^2}{m_\phi} ~,
\end{eqnarray}
which can be smaller than the decay constant without any fine-tuning.
This is the case if
\begin{eqnarray}
\label{cond}
m_\phi < H_\text{inf} < \left(\frac{8 \pi^2}{3}\right)^{1/4} \sqrt{m_\phi f_\phi} ~.
\end{eqnarray}
Note that, once the above condition is satisfied, the axion potential is well approximated by the quadratic
one as (\ref{pot}).
The   energy density of the axion at the onset of oscillations is given by
\begin{eqnarray}\label{rhoini}
\rho_{\phi,{\rm ini}} \,\simeq\, \frac{3}{16\pi^2}  H_\text{inf}^4 ~,
\end{eqnarray}
which solely depends on $\Hinf$. 
Plugging Eq.\,\eqref{phi_ini} into Eq.\,\eqref{abundance}, we obtain 
\begin{eqnarray}\label{Y3}
\Omega^{\rm stable}_{\phi} h^2  \,\simeq\,
\begin{cases}
\displaystyle
\,1.3\times 10^{-20} % \,\text{GeV} 
\bigg(\frac{g_{\star,\text{osc}}}{106.75}\bigg)^{-1/4}
\bigg(\frac{m_\phi}{0.1\,\text{GeV}}\bigg)^{-3/2}
\bigg(\frac{H_\text{inf}}{\text{GeV}}\bigg)^{4}
& {\rm for~\,}\Gamma_\text{inf} \,>\,  m_\phi
\\[0.5cm]
\displaystyle
\,1.0\times 10^{-30} % \,\text{GeV} 
\bigg(\frac{T_\text{RH}}{20\,\text{MeV}}\bigg) 
\bigg(\frac{m_\phi}{0.1\,\text{GeV}}\bigg)^{-2} 
\bigg(\frac{H_\text{inf}}{\text{GeV}}\bigg)^4
& {\rm for~\,}\Gamma_\text{inf} \,<\, m_\phi
\end{cases} 
~,
\end{eqnarray}
which is much smaller than the naive estimates (\ref{abundance}).

\begin{figure}[t!]
\begin{center}  
\includegraphics[width=145mm]{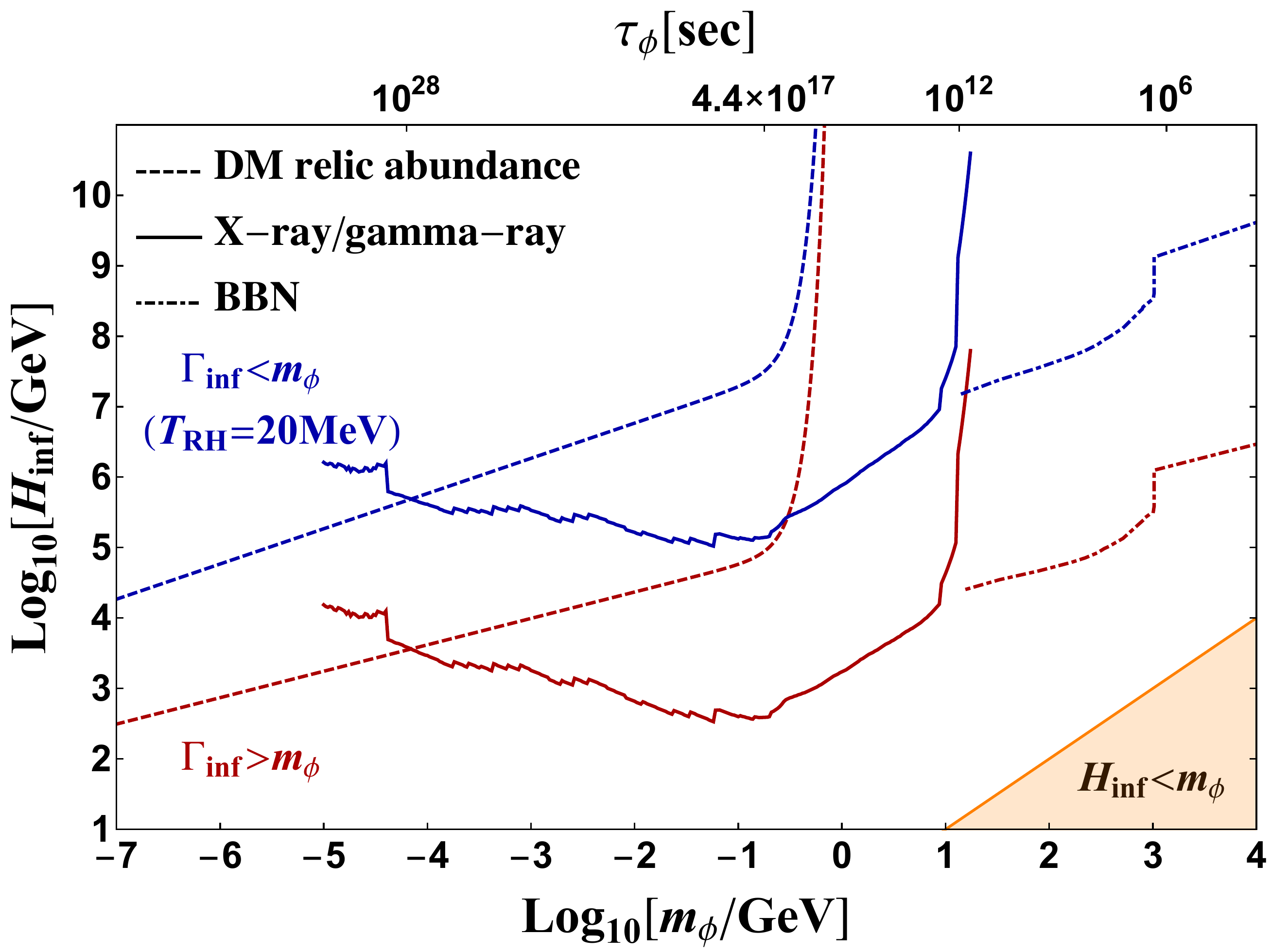}
\end{center}
\caption{The  astrophysical and cosmological upper bounds on $H_\text{inf}$ as a function of $m_\phi$, for solving the
cosmological moduli problem due to the axion. 
Here we have fixed $f_\phi = 10^{16} \,\text{GeV}$.}
\label{fig:Result}
\end{figure}

\begin{figure}[t!]
\begin{center}  
\includegraphics[width=145mm]{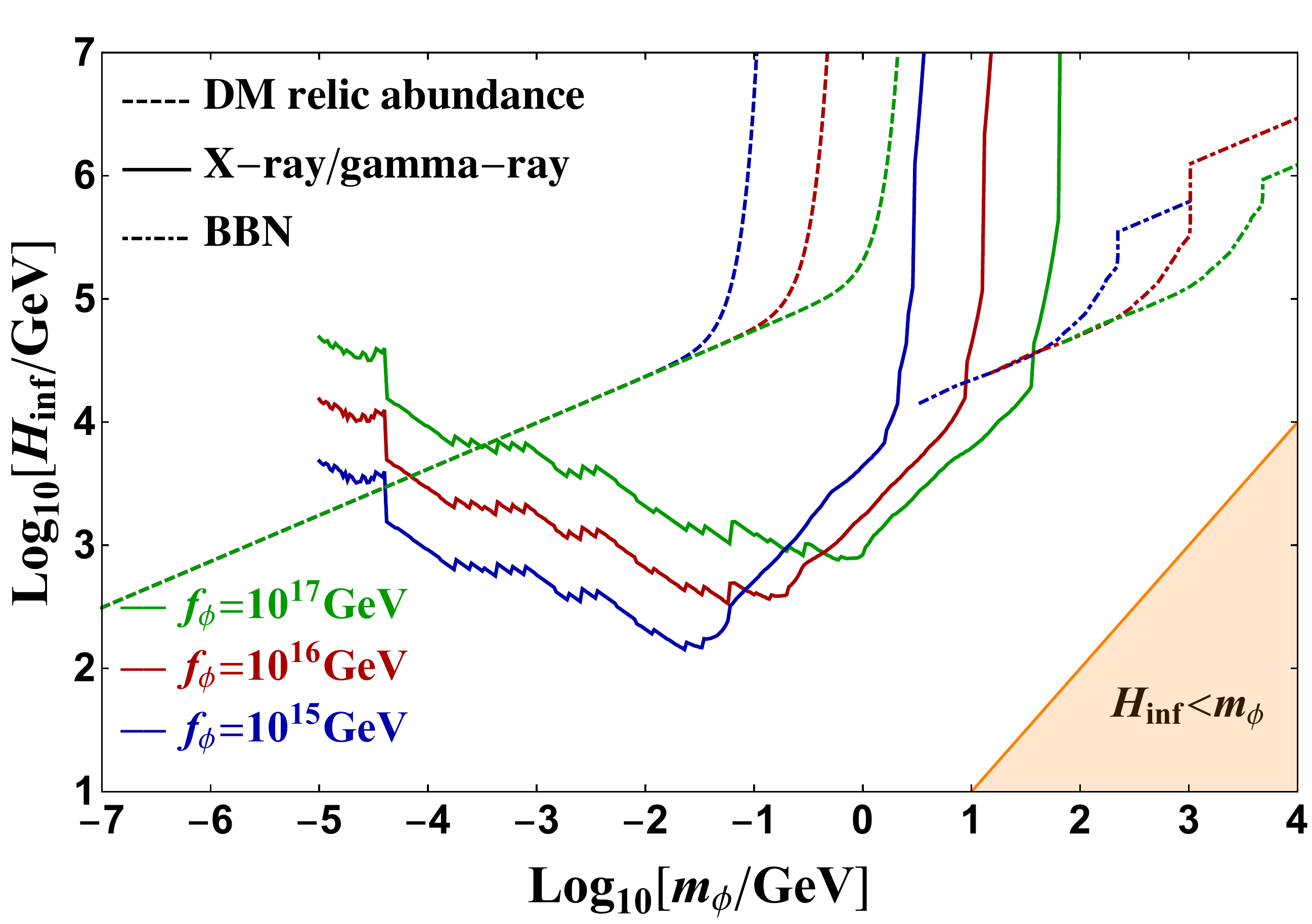}
\end{center}
\caption{Same as Fig.\,\ref{fig:Result}, but with different $f_\phi$.  The blue, red, and green lines correspond to $f_\phi =10^{15} \,\text{GeV}, 10^{16} \,\text{GeV},$ and $10^{17} \,\text{GeV}$, respectively.   $\Gamma_{\rm inf}> m_\phi$ is assumed for all the cases.}
\label{fig:Result2}
\end{figure}

\begin{figure}[t!]
\begin{center}  
\includegraphics[width=145mm]{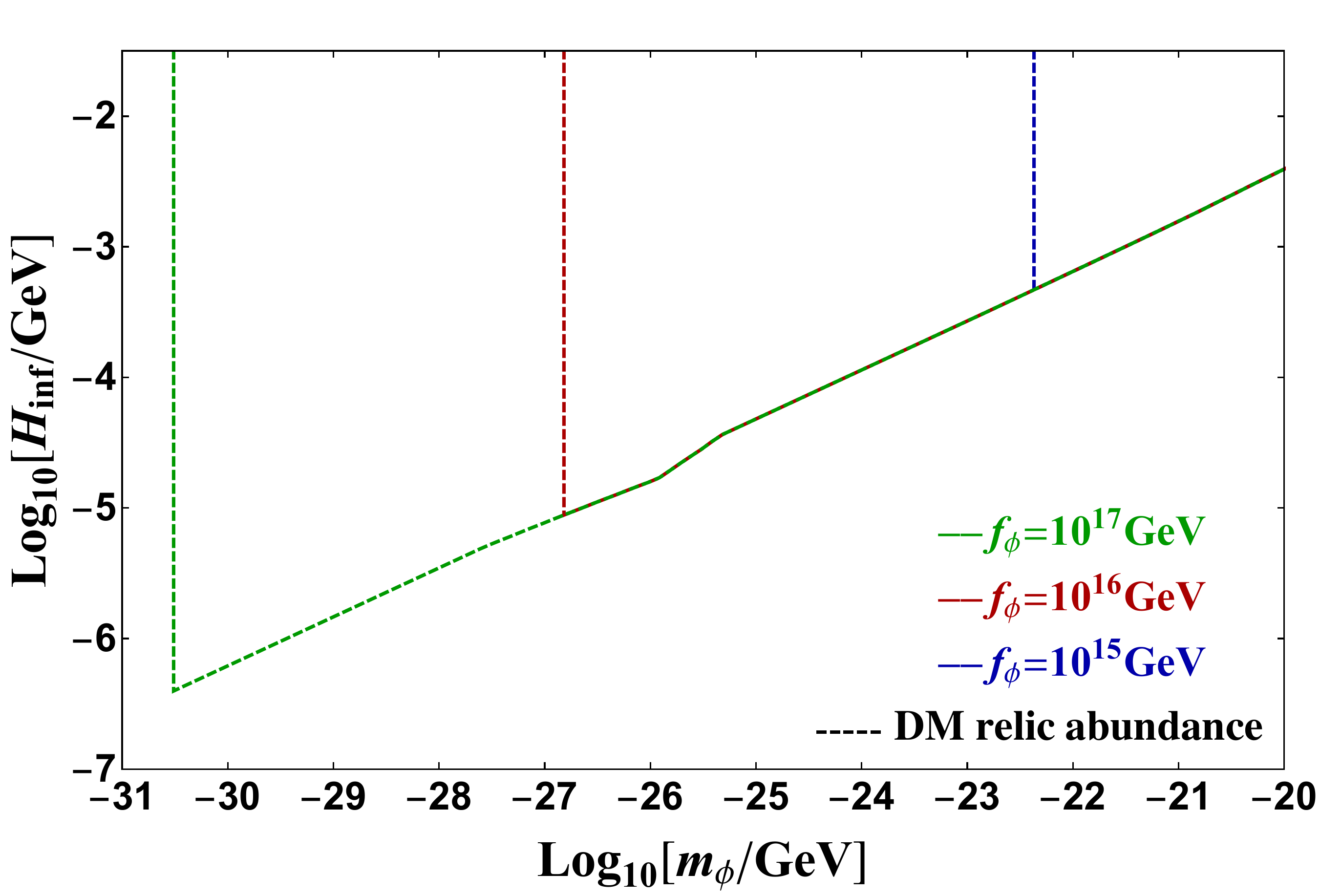}
\end{center}
\caption{Same as Fig.\,\ref{fig:Result2} but for lower axion masses. The correct DM abundance \eqref{abundance} 
is realized with $\phi_{\rm ini}=f_\phi$ on the vertical lines for different values of $f_\phi$.}
\label{fig:Result3}
\end{figure}

In order to actually satisfy the constraints given in Sec.\,\ref{subsec:2}, the Hubble parameter during inflation
is bounded above, which is shown in Fig.\,\ref{fig:Result} as a function of the axion mass. In this plot, the red (blue) 
line corresponds to the case where the axion starts to oscillate after (before) the reheating. 
The dashed, solid, and dotted-dashed lines are the bounds coming from the observed  DM abundance \eqref{DMbound}, the X-ray and gamma-ray observations~\cite{Essig:2013goa}, and the BBN~\cite{Kawasaki:2017bqm}, respectively. We also show
the region with $H_\text{inf} < m_\phi$, where the moduli problem was considered to be absent. 
By considering the BD distribution of the axion, the moduli problem for the axion is significantly relaxed, and the whole region below the red or blue lines is now allowed. 
In Fig.\,\ref{fig:Result2}, we also show the same constraints for $f_\phi=10^{15} \,$GeV (blue lines) and $10^{17}\,$GeV (green lines), for the case where the axion begins to oscillate after the reheating. For comparison, the case of $f_\phi=10^{16}\,$GeV is also shown as red lines. 
As shown in Fig.\,\ref{fig:Result3}, one can extrapolate the bound on the axion abundance (dashed lines)
toward lighter axion masses until the axion abundance becomes equal to the DM abundance for $\phi_{\rm ini}^{\rm (BD)} = f_\phi$.
For $f_\phi = 10^{15}\,, 10^{16}, 10^{17}$\,GeV, this corresponds to $m_\phi \sim 10^{-13}\,,10^{-18},10^{-22}$\,eV and $H_{\rm inf} \sim 100\,,10, 0.1$\,keV, respectively.

Lastly, let us comment on the  assumption about the axion potential. We have assumed that the axion 
potential is present during inflation and it remains unchanged after inflation. This is the case if
all the relevant dynamical scales for the axion potential are much higher than the Hubble scale 
during inflation. Even if the axion potential receives some corrections after inflation and 
the potential minimum is shifted by some amount, our mechanism
still relaxes the moduli problem as long as the shift (modulo $2\pi f_\phi$) is smaller than the typical oscillation amplitude
at that time.\footnote{Even if the location of the axion potential minimum changes by a nonzero integer times $2 \pi f_\phi$, no extra coherent oscillations are induced as long as the shift takes place in either a strongly damped or adiabatic regime.} Also, it is possible that the axion potential existed during inflation but disappears after inflation
as the hidden sector responsible for the axion potential is reheated by the inflaton decay. In this case,
the axion potential is considered to reappear as the temperature goes down due to the cosmic expansion. 
The axion abundance can be still suppressed by our mechanism in this case as long as the potential minimum
is not shifted or the shift is sufficiently small. The latter case is similar to the scenario of the QCD axion
considered in Refs.~\cite{Graham:2018jyp,Guth:2018hsa}. Also, as emphasized in Ref.~\cite{Guth:2018hsa},
the axion should not have a large mixing with the inflaton field that induces a large shift of the potential minimum (modulo $2\pi f_\phi$).\footnote{If the inflaton is CP-even and CP is a good symmetry, 
one can forbid the mixing. Also, if the inflaton enjoys the discrete shift symmetry, i.e. the inflaton is another axion, the shift of the minimum can be an integer times $2\pi f_\phi$. (c.f. Refs.~\cite{Daido:2017wwb,Daido:2017tbr, Takahashi:2019qmh}.)  }

\subsection{Generalization to multiple axions}
Before closing this section, let us briefly discuss a more general case with multiple axion fields, $\phi_i$, 
which in general have mixings as
\begin{eqnarray}
\label{geneH}
V(\phi_j) \,=\,  \sum^{N_S}_{a=1}{\Lambda_a^4 \left[1-\cos\left(\sum_{j=1}^{N_A} {c_a}^j\frac{\phi_j}{f_j}+\theta_{a} \right)\right]} ~,
\end{eqnarray}
where $N_S$ and $N_A$ are the numbers of the cosine terms and the axion fields, respectively,
 ${c_a}^j$ is an anomaly coefficient, and  $\theta_a$ is a CP phase. The decay constants, $f_j$,
are set to be of the same order.  We assume that the relevant dynamical scales for generating the axion potential are
much higher than the inflation scale so that the axion potential remains unchanged after inflation.
We exclude a case in which the axion potential is significantly modified after inflation or there is a phase transition
(e.g. bubble formation) of the axions.\footnote{
For instance, real parts of the moduli fields may not be at the vacuum due to the Hubble-induced masses, and then
some of the nonperturbative effects might be absent during inflation but they appear after inflation.
}
 Such multiple axions with mixings were
 discussed in a context of inflation~\cite{Higaki:2014pja,Higaki:2014mwa,Wang:2015rel,Masoumi:2016eqo} or DM ~\cite{Daido:2016tsj} and called ``the axion landscape".

Let us assume that inflation is driven by another sector for simplicity, although it is possible to implement the slow-roll or eternal
inflation in the context of axion landscape~\cite{Higaki:2014pja,Higaki:2014mwa,Wang:2015rel,Masoumi:2016eqo}, 
and the following argument can be straightforwardly extended to such a case.
If $N_S\geq N_A\geq 1$, all the axions generically have nonzero masses. Suppose that
they stay near one of the local minima for a long time during inflation.\footnote{Even if they are allowed to 
tunnel into a lower energy minimum, the following argument remains unchanged as long as the bubbles do not 
percolate and they occupy only a small fraction of the space-time. In this case, all the local minima will be populated in the end
and the axion distribution will be (approximately) given by the BD distribution around each minimum with a different axion
mass and the Hubble parameter.
 }
Then, the axion potential can be well approximated by
the  quadratic terms around the local minimum after a proper redefinition of the axion fields, 
\begin{eqnarray}
V(\hat{\phi}_j) \,\simeq\, \sum_{i=1}^{N_A} \frac{1}{2} m^2_j {\hat{\phi}_j^2} ~,
\end{eqnarray}
where $\hat{\phi}_j$ is the $j$-th mass eigenstate, and $m_j\,(>\,$0) is the corresponding mass eigenvalue satisfying
$m^2_1 \leq m^2_2 \leq \cdots \leq m^2_{N_A}$. Then, each mass eigenstate, $\hat{\phi}_j$, follows the BD distribution 
during inflation, and it starts to oscillate about the origin when the Hubble parameter becomes comparable to the mass after inflation.
Its initial abundance is given by  Eq.\,\eqref{rhoini} if each mass, $m_j$, satisfies the condition~(\ref{cond}) with $f_j \sim f_\phi$.

If one (or more) of $\Lambda_a$ is much smaller than the others, 
there might be a very light axion, $\phi_L$. This is indeed the case if $N_S = N_A$. Such light axion
may fluctuate over a wide field range larger than the decay constant. Even in this case, 
the abundances of heavier axions satisfying (\ref{cond}) are not modified significantly. This is because the light axion
$\phi_L$ is almost decoupled from the heavy axions, and its mixing angles are suppressed by the mass squared ratio. 
Thus, one can separately discuss the heavy and light axions.

One exception is the inflaton. During inflation, the inflaton
necessarily deviates from the low-energy minimum. And so, if the inflaton has a sizable mixing with the axion,
 the axion potential minimum is generically shifted by  a large amount, which could spoil the mechanism~\cite{Guth:2018hsa}. 
 As noted in footnote\,3, the exception is the case when the inflaton is another axion. In this case the shift can be small (modulo $2\pi f_\phi$).
% Therefore, the sizable mixing between the axion and inflaton should be absent for our mechanism to work.

%%%%%%%%%%%%%%%%%%%%%%%%%%%%%%%%%%%%%%%%%%%%%%%%%%%%%%%%%%%%%

\section{Discussion and Conclusions}\label{sec:4}
We have shown that if the Hubble parameter during inflation is sufficiently low, 
 the cosmological abundance of string axions with $f_\phi={\cal O}(10^{15-17}) \,\text{GeV}$ 
is so small that it satisfies all the astrophysical and cosmological constraints.
In addition, if the gravitino and thus the real parts of the moduli multiplets are much heavier 
than $H_{\rm inf}$,  the abundance of the real parts of the moduli multiplets are also 
highly suppressed. In an extreme case with $H_{\rm inf}\lesssim {\cal O}{(0.1)} \,\text{keV}$,
there is no cosmological moduli problem for any  axion mass.
A natural question is, then, if we can have successful reheating and inflation model building at such low scales. 
In fact, in the ALP inflation~\cite{Takahashi:2019qmh} or ALP miracle scenario~\cite{Daido:2017wwb,Daido:2017tbr} where an axion-like particle plays the role of the inflaton,  the typical Hubble scale can be extremely low, $H_{\rm inf}< {\cal O}(1) \,\text{eV}$, while successful inflation is possible through a combination of perturbative decays and thermal dissipation effects.\footnote{
The inflation model with $H_{\rm inf}$ around or below the QCD scale was studied in Ref.~\cite{Guth:2018hsa},
where the reheating proceeds through a simple perturbative decay of the inflaton. 
}

So far, we have focused on the string axion. 
In fact, our mechanism can be applied to any light scalars such as a Nambu-Goldstone (NG) boson.
One example is  a non-linear sigma model coupled to supergravity, which naturally accommodates three 
families of the ordinary quarks and leptons~\cite{Kugo:1983ai, Yanagida:1985jc}.
The K\"{a}hler potential for the NG multiplets, $\Phi$, is given in terms of a real function $\kappa(\Phi, \Phi^\dag)$ 
which transforms under the spontaneously broken symmetry as 
\begin{eqnarray}
\kappa\big(\Phi, \Phi^\dagger\big) 
\rightarrow \kappa\big(\Phi, \Phi^\dagger\big) + f(\Phi) + f^\dagger(\Phi) ~,
\end{eqnarray}
where $f(\Phi)$ is a chiral function of the NG multiplets. This, however, does not leave the Lagrangian 
invariant in supergravity, and implies that we need a singlet multiplet $X$ transforming as $X\rightarrow X-f{(\Phi)}$ to cancel the shift~\cite{Komargodski:2010rb, Kugo:2010fs}. The resultant K\"{a}hler potential takes the form of
\begin{eqnarray}
{\cal K}\,=\, 
{\cal F}\Big[\kappa\big(\Phi, \Phi^\dagger\big)+X+{X}^{\dagger}\Big] ~,
\end{eqnarray} 
which has a shift symmetry of Eq.\,\eqref{shift}. 
 If the quarks and leptons are in pseudo-NG multiplets at a certain energy scale, 
 squarks and sleptons in the first two generations can be around or lighter than TeVs but stops can be as heavy as ${\cal O}(10)\,$TeV~\cite{Yanagida:2016kag} due to the so-called Higgs mediation~\cite{Yamaguchi:2016oqz, Yin:2016shg,Yanagida:2018eho}. The gravitino and the real moduli component $r$ are around ${\cal O}(100)\,$TeV, but $\phi$ becomes much lighter 
 due to the shift symmetry. The mechanism alleviates the astrophysical and cosmological constraints induced by $\phi$ (and those for $r$ is also alleviated if $H_{\rm inf}\ll {\cal O}(100)\,$TeV).

Our mechanism to suppress the moduli abundance by low-scale
inflation has an advantage over a late-time entropy production by e.g. thermal inflation,
because the reheating temperature can be higher which makes many baryogenesis 
scenarios feasible.   For instance, when $T_\text{RH} \gtrsim 4\times 10^8\,$GeV, thermal leptogenesis 
is possible if one of the right-handed neutrinos is so light to be thermally produced~\cite{Fukugita:1986hr}
(see also Refs.~\cite{Buchmuller:2005eh, Davidson:2008bu} for reviews).
Even if all the right-handed neutrinos are heavy and decoupled, leptogenesis via active neutrino oscillations is 
still possible if the inflaton dominantly decays into the active neutrinos~\cite{Hamada:2016oft, Hamada:2018epb}.

Now let us turn to the QCD axion. The QCD axion with the decay constant $f_a\gg 10^{12}\,$GeV is known to be 
overproduced unless the initial misalignment angle is fine-tuned~\cite{Preskill:1991kd,Dine:1982ah,Abbott:1982af}.
The abundance is suppressed for $H_{\inf}\lesssim {\cal O}(100)\,$MeV~\cite{Graham:2018jyp,Guth:2018hsa}, but
remains unsuppressed, otherwise. One way to enlarge the allowed parameter space is to make the QCD axion
heavier during inflation by making the QCD scale higher~\cite{Dvali:1995ce, Banks:1996ea, Jeong:2013xta,Co:2018phi}.
During inflation the Higgs field may acquire an expectation value, $\tilde{v}$, much larger than the weak scale, 
e.g. due to the Hubble-induced mass, or it might be trapped at the false vacuum. Then,
all the quarks obtain masses of order $\tilde{v}$ times its Yukawa coupling. 
As a result, the QCD confines at a scale, $\tilde{\Lambda}_{\rm QCD}$, much larger than $\Lambda_{\rm QCD}$~\cite{Dvali:1995ce,
Banks:1996ea, Jeong:2013xta},
\begin{eqnarray}
\tilde{\Lambda}_{\rm QCD}
\,\simeq\,
\begin{cases}
\displaystyle
10^6\,{\rm GeV} \left({\tilde{v} \over M_{\rm pl}}\right)^{4/11}
& {\rm for ~SM}
\\[0.5cm]
\displaystyle
10^7\, {\rm GeV}  \left({\tilde{v}\over 10^{16}\,{\rm GeV} }\right)^{2/3}
& {\rm for ~MSSM}
\end{cases} 
~.
\end{eqnarray}
If there are extra quarks, the QCD scale becomes even larger. For instance, there might be 
vector-like quarks coupled to the flat direction including the Higgs through higher dimensional operator~\cite{Jeong:2013xta}. 
In the MSSM case, the Hubble-induced mass easily drives the Higgs field value to be large in the flat-direction of the potential, 
but one needs to make sure that the CP-conserving minimum is not changed during inflation and at the vacuum~\cite{Choi:1996fs},
which generically necessitates additional assumptions on the set-up. A simple assumption is the minimal flavor and CP violation, where the CP phase, as well as the flavor violation of the soft parameters, originates from the CKM matrix.
In the SM case, there is no extra CP phase, but one has to have the Higgs potential energy in the false vacuum 
so small that the inflation scale does not exceed the effective QCD scale during inflation.\footnote{
The topological Higgs inflation does not work because of too high energy scale~\cite{Hamada:2014raa}.
} This may require the tuning of the
higher dimensional Higgs couplings that uplift the potential or a special value of the top quark mass that leads to the (almost)
degenerate two vacua~\cite{Froggatt:1995rt}. An interesting possibility is that eternal inflation is driven by the SM Higgs potential
energy in the false vacuum. Then, the effective QCD scale is higher than the ordinary case, and the QCD axion acquires a heavy 
mass. Their initial oscillation amplitude is determined by the BD distribution, suppressing the QCD axion abundance for a broader
range of the inflation scale. The eternal Higgs inflation ends through the tunneling of the Higgs field to the current vacuum.
The SM sector is considered to be thermalized by the latent heat through preheating, 
and afterwards the slow-roll inflation should take place in the pocket Universe to generate density perturbations 
and reheat the SM sector again. 
This is an interesting possibility in which the SM Higgs sector drives the eternal inflation and
at the same time increases the effective QCD scale, broadening the viable parameter space. 
Note that, for this mechanism to work, one needs to make sure that the Peccei-Quinn (PQ) symmetry is not restored. 
For instance,  if the scale of the eternal inflation is $H_{\rm inf}\simeq 10^6\,$GeV, and if all the energy goes to the SM radiation inside a bubble
after the tunneling, the reheating temperature will be of order $10^{12}$\,GeV. So, in this case,  the PQ scale should be 
greater than ${\cal O}(10^{12})\,$GeV in order to avoid the PQ symmetry restoration. 
A further study is warranted.

In this paper, we have shown that astrophysical and cosmological constraints on 
string axions can be significantly alleviated with low-scale inflation. 
The string axions can stay around the potential minima even for $H_{\rm inf}$  much greater than the axion 
masses if the axion field reaches the BD distribution peaked at the potential minimum.
 This is the case when the inflation lasts long enough and the minima do not change much during and after the inflation.
We have found that the cosmological abundance of the axion is significantly suppressed compared to naive estimates. 
As a result, the cosmological moduli problem induced by the string axions is significantly relaxed by the low-scale inflation, 
and astrophysical and cosmological bounds are satisfied 
over a wide range of the mass without any fine-tuning of the initial misalignment angles.

\section*{Acknowledgments}
This work is supported by GP-PU Program, Tohoku University (S.Y.H), by JSPS KAKENHI Grant Numbers JP15H05889 (F.T.), JP15K21733 (F.T.),  JP17H02875 (F.T.), JP17H02878(F.T.), by NRF Strategic Research Program NRF-2017R1E1A1A01072736 (W.Y.), and by World Premier International Research Center Initiative (WPI Initiative), MEXT, Japan.

\end{document}